\documentclass[aps,prl,superscriptaddress,showpacs,amsmath,amssymb,amsfonts,altaffillsymbol,twocolumn]{revtex4-1}

\usepackage{graphicx}% Include figure files
\usepackage{dcolumn}% Align table columns on decimal point
\usepackage{txfonts}
\newcommand{\bvec}[1]{\mbox{\boldmath $#1$}}
\graphicspath{{./figures/}}

\begin{document}

\title{
Unveiling Dimensionality Dependence of Glassy Dynamics:  \\
2D Infinite Fluctuation Eclipses Inherent Structural Relaxation
}
\author{Hayato Shiba}
\email[E-mail: ]{shiba@imr.tohoku.ac.jp}
\affiliation{Institute for Solid State Physics, University of Tokyo, Chiba 277-8581, Japan}
\affiliation{Institute for Materials Research, Tohoku University, Sendai 980-8577, Japan}

\author{Yasunori Yamada}
\altaffiliation[Present Address: ]{Beijing Computational Science Research Center, Beijing 100193, People's Republic of China}
\affiliation{Institute for Materials Research, Tohoku University, Sendai 980-8577, Japan}

\author{Takeshi Kawasaki}
\affiliation{Department of Physics, Nagoya University, Nagoya 464-8602, Japan}
\author{Kang Kim}
\affiliation{Department of Physics, Niigata University, Niigata 950-2181, Japan}
\affiliation{ Division of Chemical Engineering, Graduate School of Engineering Science,
Osaka University, Osaka 560-8531, Japan}

\date{\today}% It is always \today, today,
             % but any date may be explicitly specified

\begin{abstract}
By using large-scale molecular dynamics simulations, the dynamics of
two-dimensional (2D) supercooled liquids turns out to be dependent on
the system size, while the size dependence is not pronounced in three dimensional (3D) systems.
It is demonstrated that the strong system-size effect in 2D amorphous systems
originates from the enhanced fluctuations at long wavelengths,
which are similar to those of 2D crystal phonons.
This observation is further supported by the frequency dependence
of the vibrational density of states, consisting of
the Debye approximation in the low-wavenumber-limit.
However, the system-size effect in the intermediate scattering
function becomes negligible when the length scale is larger than the vibrational amplitude.
This suggests that the finite-size effect in a 2D system is transient and also that
the structural relaxation itself is not fundamentally different from that
in a 3D system. In fact, the dynamic correlation lengths estimated
from the bond-breakage function, which do not suffer from those
enhanced fluctuations, are not size dependent in either 2D or 3D systems.
\end{abstract}


\maketitle
Dimensionality plays a key role in the physics
of solids and liquids -- from high to low dimensions -- and
fluctuation shows up differently, as typically observed in phase transitions~\cite{Mermin1966,Kosterlitz1973}.
Indeed, two-dimensional (2D) systems often exhibit enhanced fluctuations,
leading to various anomalies that are not experienced in three-dimensional (3D) systems.
The melting of a 2D solid is a marked
example~\cite{Halperin1978,Weber1995,Strandburg1988,Shiba2009,Bernard2011,Deutschlander2014,Mazars2015},
where the long-wavelength structural correlation is induced by thermal fluctuations sthat span an infinite length.
For the glass transition from supercooled liquids to amorphous solids, 
the dimensionality dependence of the fluctuation has become an issue only recently.
Gigantic fluctuation in 2D supercooled liquids has been observed
that is far stronger than that in their 3D counterparts~\cite{Shiba2012,Flenner2015,Vivek2016}.
The aim of this Letter is to elucidate the similarity of this fluctuation to that in
crystals~\cite{KeimCommentary}, and also to investigate the heterogeneous dynamics in both 2D and 3D systems.

For a crystalline solid of monodisperse particle assemblies, the mean-squared thermal displacement (MSTD) is
given by using the vibrational density of state (VDOS) $g(\omega )$ as a function of angular frequency $\omega$  as
\begin{equation}
\langle |\bvec{u}|^2 \rangle = \frac{dk_BT}{m} \int \frac{g(\omega )}{\omega^2}\ d\omega, \label{eq:vdos}
\end{equation}
where $m$ the particle mass, $d$ the spatial dimension, and $(k_BT)^{-1}$ the inverse temperature.
Under the Debye approximation for
the VDOS of acoustic plane waves,
$g (\omega )$ becomes proportional to $\omega^{d-1}$~\cite{Ashcroft1976}.
It leads to divergence of the integral in 2D systems owing to the low-frequency acoustic waves,
while it converges in 3D systems.
As a result, the long-range translation order is prohibited in 2D systems~\cite{Jancovici1967,Imry1971}.
Integration of Eq. (\ref{eq:vdos}) over $\omega \ge 2\pi c / L$ provides us 
with its dependence on the linear system size $L$ as 
\begin{equation}
\langle | \bvec{u}|^2 \rangle
\sim \frac{k_BT}{2\pi}\left( \frac{1}{\mu} + \frac{1}{K+\mu } \right) \ln \left( \frac{L}{\sigma_0}\right),
\label{eq:elastic}
\end{equation}
where $\mu$ and $K$ are shear and bulk moduli, $\sigma_0$ is the particle radius,
and $c$ is the velocity of sound.
Such fluctuation is the source of the size-dependent behavior
of 2D solids undergoing melting~\cite{Weber1995,Shiba2009,Bernard2011,Mazars2015}.

In amorphous solids, $g(\omega )$
is known to behave quite differently than it does in crystals.
Simulations~\cite{Taraskin1997,Schirmacher1998,Xu2007,Shintani2008,Monaco2009,Tan2012,Ruocco2013} 
and experiments~\cite{Duval1990,Greaves2007,2009Monaco2,Yodh2010,Chumakov2011}
indicate an abundance of acoustic excitations in amorphous solids,
exhibiting the so-called boson peak. 
For frequencies far lower than the boson peak, the Debye model can be supposed because
the microscopic details are irrelevant to long-wavelength modes~\cite{2009Monaco2,Klix2015}.
Several attempts have been made to verify the Debye model description on the microscopic basis
by using molecular simulations~\cite{Leonforte2005,Monaco2009}, but no conclusive
simulation data have been provided on the asymptotic behaviors at low frequencies.
Therefore, it is still an open issue how these low-frequency vibration modes
affect the dimensionality dependence of thermal fluctuations. 

In this study, we address the dimensionality dependence
of the low-frequency  thermal vibrations and dynamics.
The 2D binary 50:50 soft-core (2D SC)~\cite{Yamamoto1997,Yamamoto1998a}
and 3D binary 80:20 Kob--Andersen-type Lennard--Jones (3D KALJ)~\cite{Kob1995}
potentials are used for our simulation.
The standard Newtonian dynamics simulation is performed for supercooled states. 
The simulation results are presented in terms of the reduced units~\cite{suppl}. 
Simulations have also been performed
for 2D KALJ and 3D SC with smaller system sizes, but the results do not qualitatively differ.

\begin{figure}
\includegraphics[width=\linewidth]{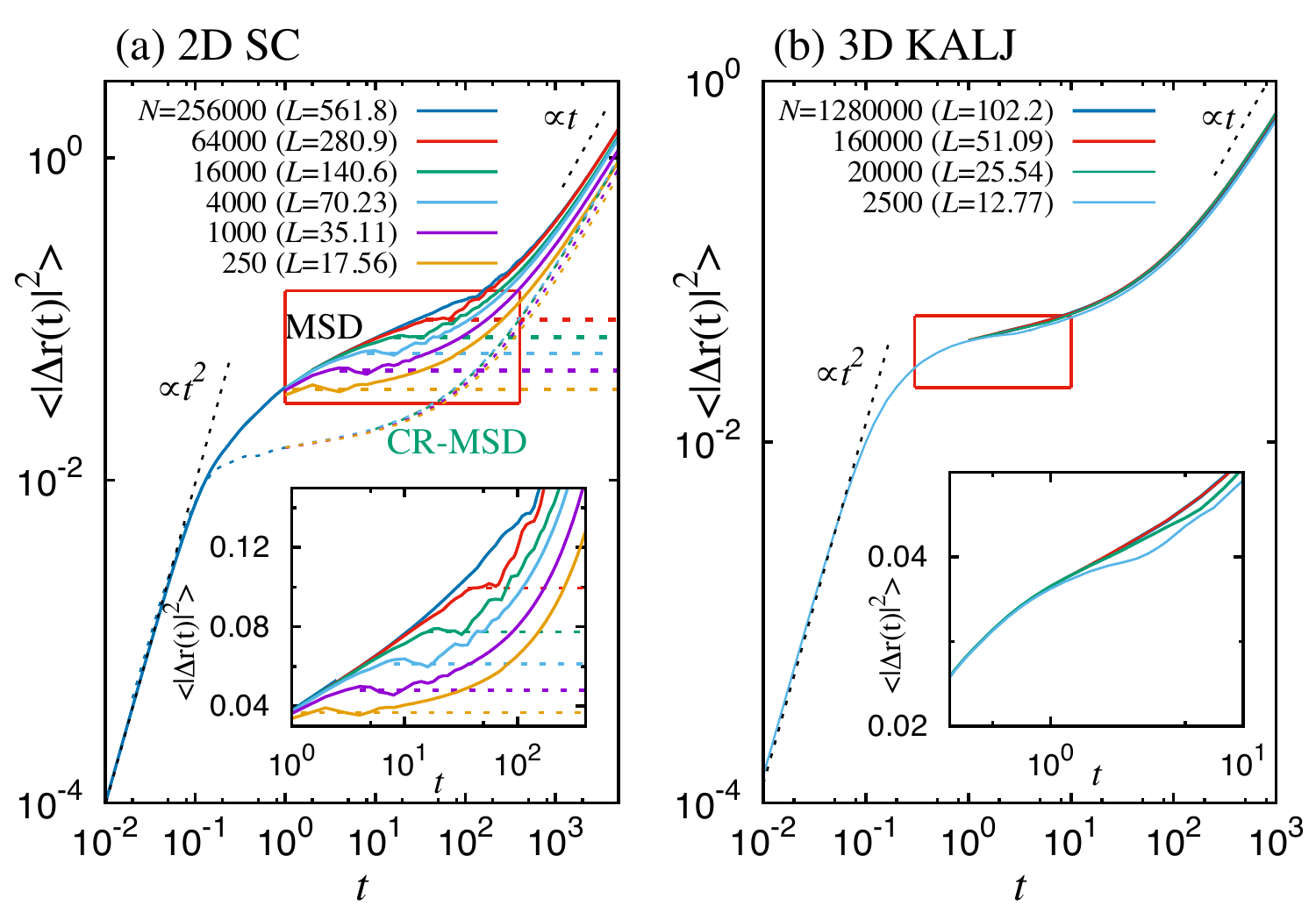}
\caption{\label{fig:MSD_Ndep}
Solid lines indicate the MSD $\langle |\Delta\bvec{r}(t)|^2\rangle$ of
(a) 2D SC at $T=0.64$ and (b) 3D KALJ at $T=0.47$ for various system sizes.
In the insets, the respective plateau regions (indicated by red boxes)
are magnified. In (a), the height of the plateaus are indicated by the 
horizontal dotted lines, and the curved dotted lines indicate the CR-MSDs 
$\langle |\Delta \bvec{r}_{\rm CR} (t) |^2\rangle $ for various
system sizes $N\le 64\ 000$ with the correponding colors.
}
\end{figure}

First, we study the mean-square displacements (MSDs)
of 2D SC and 3D KALJ.
Between the short-time ballistic and long-time diffusive regimes,
there exists a plateau in each MSD, and its height can be 
estimated directly as its magnitude 
$\langle |\Delta\bvec{r}(t_p) |^2 \rangle$ at the plateau time $t=t_p$.
In Fig. \ref{fig:MSD_Ndep}, the MSDs $\langle |\Delta \bvec{r} (t)|^2\rangle = 
\langle (1/N) \sum_j |\Delta\bvec{r}_j (t) |^2 \rangle$ are plotted for (a) 2D SC  and (b) 3D KALJ,
with $\Delta \bvec{r}_j (t) = \bvec{r}_j(t) - \bvec{r}_j (0)$ the particle displacement. 
For particle numbers $N=250$, 1\ 000, 4\ 000, and 16\ 000 in 2D SC,
the plateaus are distinctly observed, as indicated by the horizontal lines in (a).
The plateau is a bit raised for $N=64\ 000$, exhibiting
a crossover to the long-time diffusive regime.
Finally, the plateau disappears for $N=256\ 000$.
Cage-relative MSDs (CR-MSDs) $\langle |\Delta\bvec{r}_{\rm CR}(t)|^2 \rangle$ ~\cite{2009Mazoyer,2011Mazoyer,2016Illing}
are also plotted for 2D SC with $N\le 64\ 000$. 
CR-MSD is defined as the averaged mean-square of the 
displacement $\Delta \bvec{r}_{i,{\rm CR}} (t)$ that is relative to the center of mass of neighboring  particles,
and its significance will be addressed later. 
The collapsed data of CR-MSD indicate absence of finite-size size effect in CR-MSD. 
In 3D KALJ in (b), the MSDs exhibit virtually no size dependence for $N\ge 2\ 500$.
 
\begin{figure}
\includegraphics[width=\linewidth]{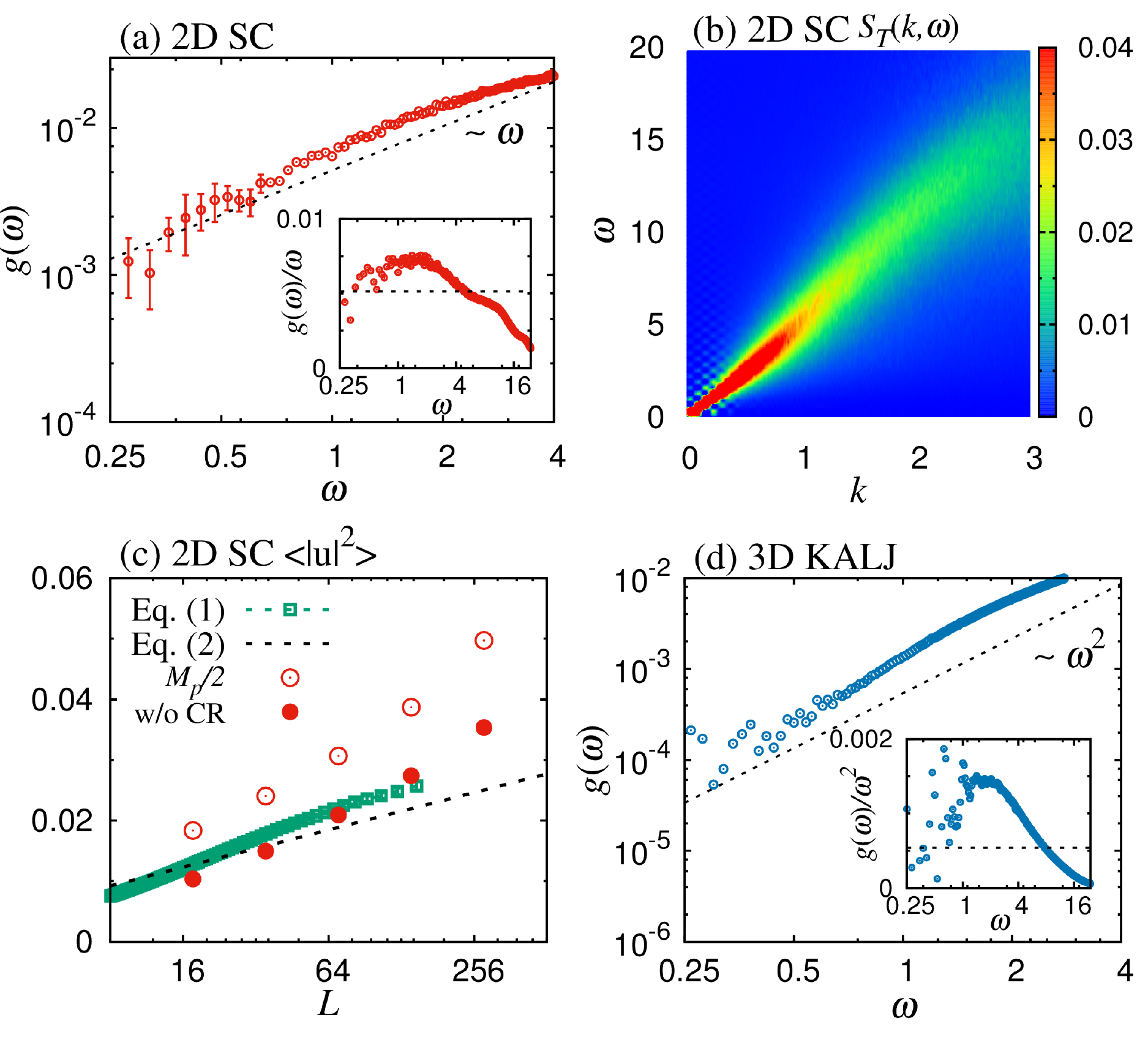}
\caption{\label{fig:dos}
(a) The VDOS $g(\omega )$ of 2D SC for $N=256\ 000$. The black dotted line
represents Debye asymptote.
Inset: Replot in the form of $g(\omega )/\omega$. 
(b) Color plot of the DSF for the transverse wave $S_T (k,\omega )$ of 2D SC,
representing the linear dispersion relationship.
(c) MSTD $\langle |\bvec{u}|^2\rangle$ 
estimated from the VDOS for 2D SC at $T$=0.64
is compared with the MSD plateaus. 
Green plots stand for the direct integration of Eq. (\ref{eq:vdos})
in the range $2\pi c_T / L \le \omega < \infty$ in (a), for which 
all the errors are inside the box symbols, 
and the black dotted line is the estimate by Eq. (\ref{eq:elastic}).
Red open circles represent a half of the MSD plateau height 
$\langle |\Delta \bvec{r}(t_p)|^2\rangle$ in Fig. \ref{fig:MSD_Ndep} (a).
The red filled points are the values after subtraction of CR-MSD, 
$\langle |\Delta \bvec{r}(t_p)|^2\rangle - \langle |\Delta \bvec{r}_{\rm CR} (t_p)|^2 \rangle$, 
which stand for squared thermal amplitude of the long-wavelength modes.
(d) The VDOS $g(\omega )$  of 3D KALJ for $N=10\ 240\ 000$.
The black dotted line represents the Debye asymptote.
Inset: $g(\omega)/ \omega^2$ is plotted for $N=160\ 000$.
}
\end{figure}

This size dependence in the MSDs is attributed to the long-wavelength acoustic sound modes,
represented in the VDOS at low frequencies comparable to 
$\omega \sim 2\pi c/L$ --  the limiting behavior in the limit of $\omega\rightarrow 0$  matters.
Although the VDOS is usually estimated by the normal mode analysis,
it becomes more difficult as the system size becomes larger, because
the eigenvector calculation of the $dN\times dN$ Hessian matrix is required.
As an alternative, the VDOS is obtained by
directly calculating the velocity correlation function (VCF)
as  $g(\omega ) = 2 (\pi dN k_BT)^{-1} \int dt\ e^{i\omega t} \sum_j
\langle m_j \bvec{v}_j (0)\cdot\bvec{v}_j(t) \rangle $~\cite{Shintani2008},
with $m_j$ and $\bvec{v}_j (t)$ representing the mass and the velocity of particle $j$.
The time development of the thermal vibration
has been simulated at low temperatures, $T=0.01$  for 2D SC and $0.008$ for 3D KALJ.
The initial particle configurations are prepared by the steepest descent method,
in order to begin the simulation from the local potential minimum. 
We find that these temperatures are low enough that the 
obtained VDOS provides the faithful description of the normal modes. 
In Fig. \ref{fig:dos} (a),
the VDOS is  shown for 2D SC with $N=256\ 000$.
In the limiting behavior $\omega \rightarrow 0$,
the VDOS clearly exhibits linear $\omega$ dependence. 
In the Debye theory of the crystalline solids~\cite{Ashcroft1976},
the VDOS can be explicitly estimated to be $g_D (\omega ) =  \omega / (2\pi n c_M^2)$,
where $n=N/L^d$ and $c_M^{-2} = ( c_L^{-2} + c_T^{-2})/2$,
with $c_L$ and $c_T$ being the longitudinal and transverse sound velocities, respectively.
The VDOS asymptotically approaches $g_D(\omega )$, assuming 
up to 1.4 times larger values in the present frequency region. 
The sound velocities are estimated by dynamic structure factors (DSFs)
$S_{L,T} (k,\omega) =  2\bar{m} (d\pi N k_BT)^{-1}  \int dt\ e^{i\omega t} \langle \bvec{j}_{L,T}(\bvec{k},0)^*
\cdot\bvec{j}_{L,T}(\bvec{k},t)\rangle$  with $\bvec{j}_{L,T} (\bvec{k},t)$ representing the longitudinal and transverse
current velocities~\cite{Ruocco2013,Mizuno2014} and $\bar{m}$ the average particle mass.
The wavenumbers $k = |\bvec{k}|$ at the peak values of the DSFs satisfy the linear dispersion relations $\omega = c_{L,T} k$,
as seen for $S_T(k,\omega)$ of 2D SC in Fig. \ref{fig:dos} (b).
The linear fit provides the estimate values of sound velocities as
$c_L$ = 11.9 and 8.6 and $c_T$ = 4.7 and 3.8 for 2D SC and 3D KALJ, respectively.

If the MSD plateau height $\langle |\Delta \bvec{r}(t_p)|^2 \rangle$ 
arises as a superposition of harmonic vibration modes,
it is equal to the squared thermal amplitude $A_p^2 = 2\langle |\bvec{u}|^2\rangle$. 
Therefore, the half of the plateau height is a value that can be directly
compared with the MSTD $\langle |\bvec{u}|^2\rangle$ estimated either 
from the VDOS data using Eq. (\ref{eq:vdos}) or from the Debye approximation in Eq. (\ref{eq:elastic}).
For both Eq. (\ref{eq:vdos}) and Eq. (\ref{eq:elastic}), 
the integration is cut off at $\omega_{\rm min} = 2\pi c_T /L$, 
so that the MSTD can be estimated as a function of the box length $L$. 
For the direct integration of $g(\omega)$ in Fig. \ref{fig:dos} (a),
the size dependence is mainly due to a small number of available
low-frequency modes ($\omega \lesssim 1$). 
On that account, the integration is performed for 
$L<148\sigma_{11}$,  
with $\sigma_{11}$ being the radius of the 1st component particles, and
all of the particle masses are replaced by its average $\bar{m}$. 
For Eq.  (\ref{eq:elastic}),  the shear and bulk moduli of 2D SC  
are  $\mu = \rho c_T^2 = 26.6 $ and $K= \rho c_L^2 - \mu = 142.1$, with $\rho = \bar{m} n$ being the mass density. 
In the comparison  given in Fig. \ref{fig:dos} (c), 
half of the MSD plateau height (the open circles) is considerably larger than the estimations.
In order to remove contributions from the local center-of-mass fluctuations, 
CR-MSD at the same time $t_p$, $\langle | \Delta \bvec{r}_{\rm CR} (t_p)|^2\rangle$, 
is further subtracted from the plateau height $\langle |\Delta \bvec{r} (t_p) |^2 \rangle$.
We find that the subtracted value agrees with the estimations made using VDOS
where the sole effect of long-wavelength motion is taken into account
after subtraction. 
This observation further attributes the cause of 2D system-size dependence to the Debye asymptote in $g (\omega )$.

Also, in 3D systems, the VDOS 
is expected to exhibit Debye asymptote behavior at low-frequencies.
This is shown for 3D KALJ with $N=10\ 240\ 000$ in Fig. \ref{fig:dos} (d),
and it is also shown over a wider range of $\omega$ for $N=160\ 000$ in its inset.
The VDOS asymptotically approaches $g_D(\omega) = \omega^2 / (2\pi^2 n c_M^3 )$
which is given by the Debye approximation with $c_M^{-3} = (c_L^{-3} + 2c_T^{-3})/3$. 
The low-frequency modes have small influences on the integration of Eq. (\ref{eq:vdos})
because $g_D(\omega ) \sim \omega^2$ rapidly goes to zero in the $\omega\rightarrow 0$ limit.
This fact fits together with the lack of finite-size effects in MSTD in 3D KALJ.
The VDOS still exhibits the values that are 1.6 to 2.6 times larger values 
in the range $0.25 \lesssim \omega \lesssim 1$. 
It remains an open question whether or not the VDOS further approaches 
the Debye asymptote at lower frequencies. 
It is notable that a previous experiment also shows a few times larger VDOS
than the Debye asymptote~\cite{2009Monaco2}.
 
\begin{figure}
\includegraphics[width=.8\linewidth]{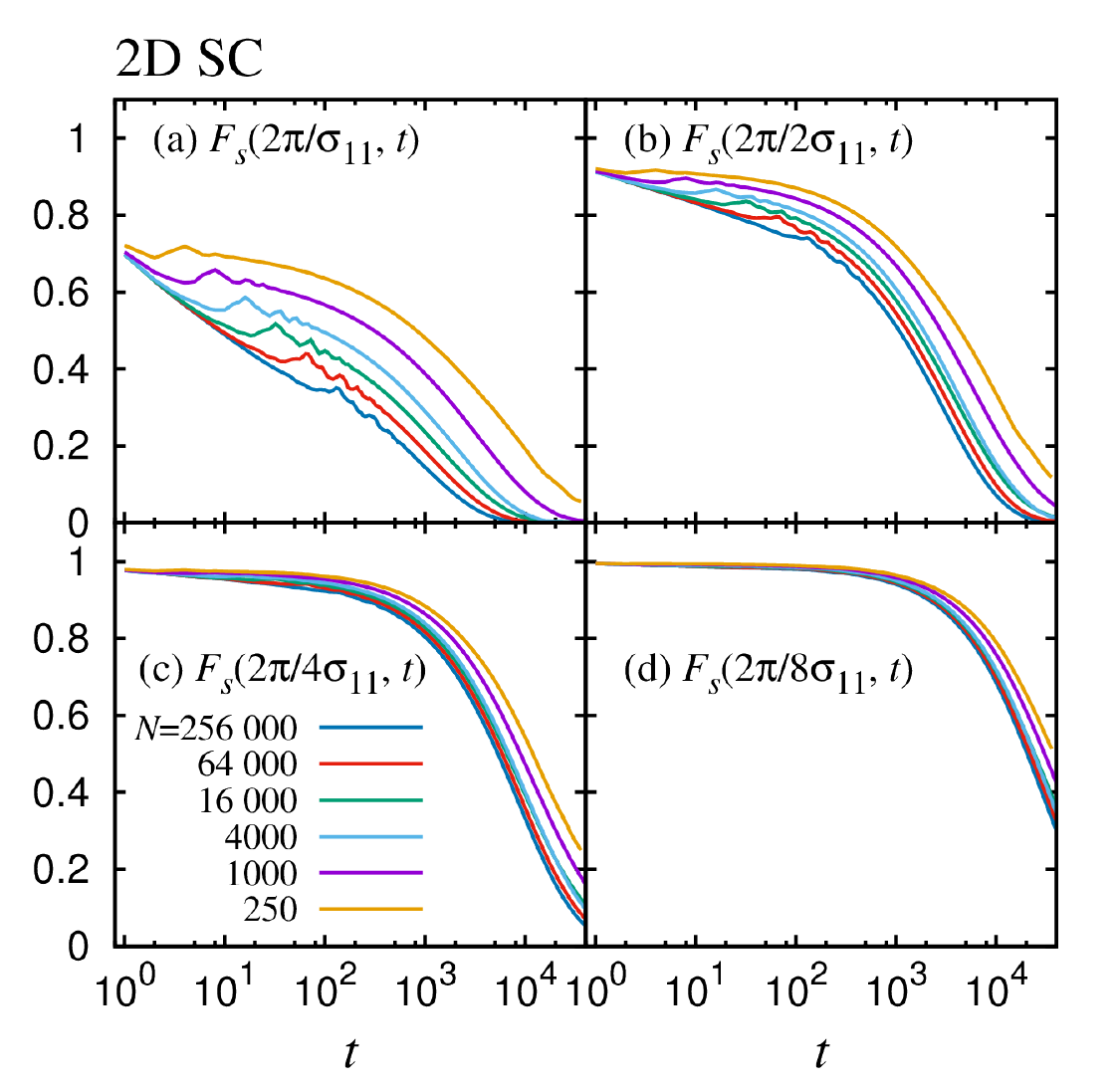}
\caption{\label{fig:Fskdep_N}
Size dependence of $F_s(k = 2\pi / \lambda,t)$ of 2D SC at $T=0.64$
is plotted, for $\lambda=$ (a) $\sigma_{11}$, (b) $2\sigma_{11}$, (c) $4\sigma_{11}$, and (d) $8\sigma_{11}$.
}
\end{figure}

Now that the source of the 2D anomaly has been revealed,
we address the resultant dimensionality dependence of the 
manner of structural relaxation,
not only by investigating the density correlation but also 
by looking into the dynamic heterogeneity (DH)~\cite{Kob1997,Yamamoto1997,Yamamoto1998a,Yamamoto1998b,Berthier2011a,
Lacevic2003,Szamel2006,Flenner2011}. 
In the literature, DH is considered to be one of the fingerprints of vitrification,
and its cause is attributed to consecutive intermittent jump motions 
of particles escaping out of cages~\cite{Donati1998,Weeks2002,Kawasaki2013a,Pastore2014,Ciamarra2016}.
To begin with, the system-size dependence of the 
self-part of the intermediate scattering function
$F_s ( k,t ) =  (1/N) \sum_j \exp [i\bvec{k}\cdot
(\bvec{r}_j (t) -\bvec{r}_j (0)  )]$ is shown for 2D SC in Fig. \ref{fig:Fskdep_N}. 
In addition to the standard wave-number $k =  2\pi /\sigma_{11}$, 
three smaller wave-numbers are considered. 
$F_s (k ,t)$ relaxes faster with larger system sizes for $k = 2\pi /\sigma_{11}$, 
consistent with a previous result for 2D KALJ~\cite{Flenner2015}.
For a smaller $k$, the relaxation becomes independent of the system size.
Therefore, the particle motion on a length scale a few times larger than
$\sigma_{11}$ does not depend on the system size, while motion on the particle-size scale depends
(note that $A_p$ amounts to about $0.3\sigma_{11}$ in Fig. \ref{fig:dos} (c) ).
The finite-size effect is observed merely as a transient effect taking place at short time 
 and length scales induced by vibrations. It is backed by the periodic transient peaks due to the
sound waves traversing over the system for small values of $\lambda$.
In 3D systems, $F_s (k,t)$ (not shown) is independent of the system size, 
even for $k = 2\pi/ \sigma_{11}$.

DH is investigated in terms of the following variables that characterize the dynamics~\cite{suppl}.
One is the four-point correlation function~\cite{Lacevic2003,Szamel2006,Chandler2006,Flenner2010,Flenner2011}  
that characterizes the configuration overlap,
by the overlap function $W_j(t) = \Theta (a- |\Delta \bvec{r}_j(t)| )$
using the Heavisidie step function $\Theta$. 
The overlap function assumes a value of unity if the particles move over a distance longer than $a$ and zero otherwise.
The threshold distance $a$ is set to $0.3\sigma_{11}$ in accord with the standard choice. 
The other  bond-breakage function~\cite{Yamamoto1997,Yamamoto1998a,Kawasaki2013a,Flenner2016} 
is based on the broken-bond number $Z_j(t)$, which
 addresses the correlation time and length caused by the change in 
the local particle connectivity~\cite{Iwashita2013}.
$Z_j(t)$ starts from 0 at the initial time $t=0$ and
increases one by one as $t$ proceeds, as a pair of
bonded particles get separated from each other.
The DH can then be probed through these functions' respective dynamic susceptibilities
$\chi_4(t)$ and $\chi_B(t)$, and also the 
corresponding structure factors $S_4(k,t_4)$ and $S_B(k,t_B)$
at the respective peak times $t_4$ and $t_B$ of the susceptibilities,
which describe the wave-number dependence of the heterogeneous motions.
From both the structure factors,
the respective dynamic correlation lengths $\xi_4$ and $\xi_B$  can be estimated
by fitting with the generalized Ornstein-Zernike (OZ) function~\cite{Kim2013a}.  

\begin{figure}
\includegraphics[width=\linewidth]{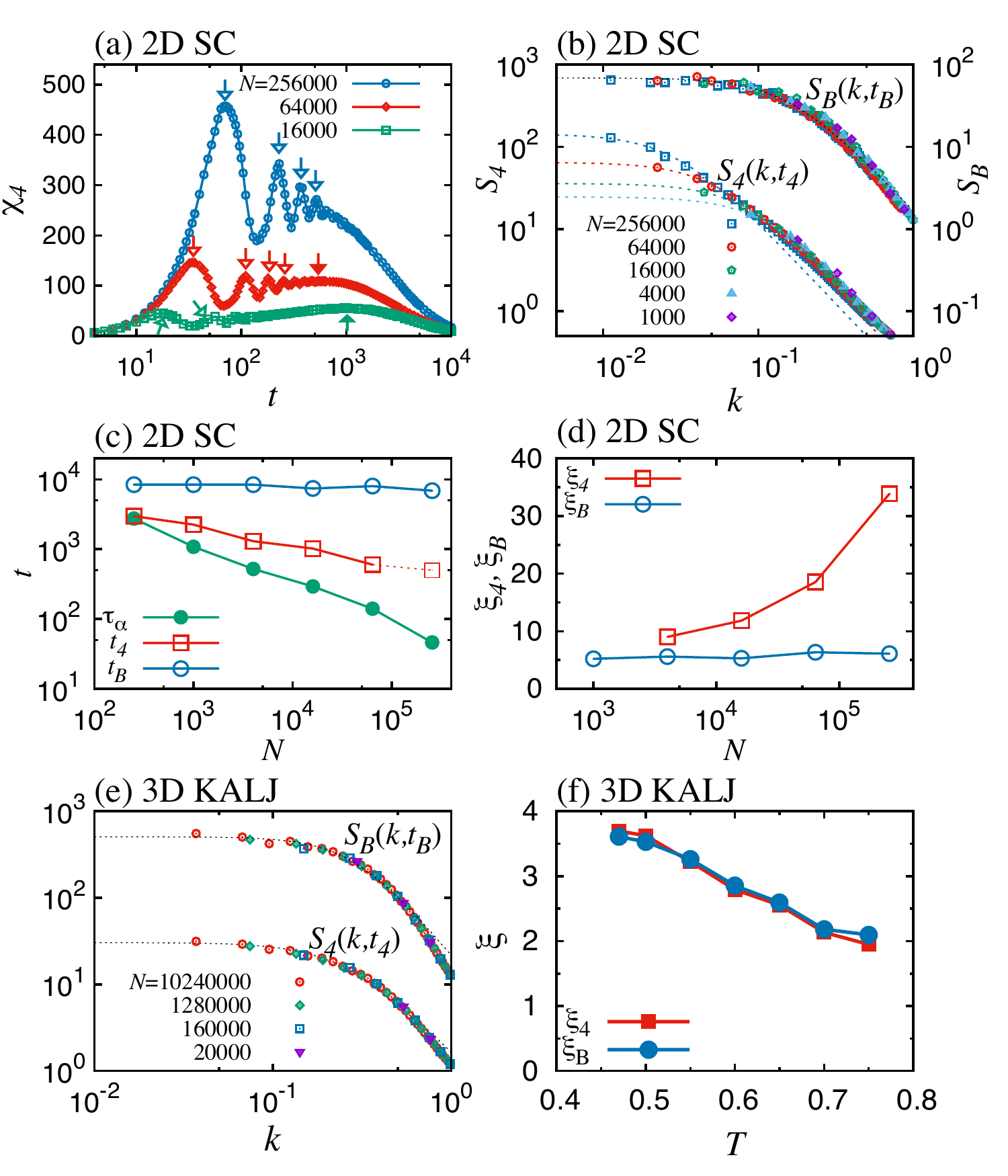}
\caption{\label{fig:DH}
(Color online)  (a) and (b) show the size dependence
of (a) $\chi_4 (t)$  and (b) $S_4 (k,t_4)$ and $S_B(k,t_B)$ 
for 2D SC at $T=0.64$.
The open arrows in (a) indicate the peaks originating in the
transverse sound modes, and filled arrows ($N\le 6 000$) indicate $t_4$.
The dotted lines in (b) represent the generalized OZ fit for each.
(c), (d) Relaxation and correlation times ($\tau_\alpha,\ t_4,\ t_B$) and
correlation lengths ($\xi_4,\ \xi_B$) of 2D SC, with $t_4$ of $N=256\ 000$ conjectured by the
extrapolation.
(e) Size dependence of $S_4(k,t_4)$ and $S_B(k,t_B)$ of 3D KALJ at $T=0.47$.
The dotted lines represent the generalized OZ fits for $N=10\ 240\ 000$.
(f) Temperature dependence of $\xi_4$ and $\xi_B$
is plotted for a 3D KALJ with $N=1\ 280\ 000$.}
\end{figure}

The results are summarized in Fig. \ref{fig:DH}.
As shown in Fig. \ref{fig:DH} (a),  $\chi_4 (t)$ exhibits peaks for large system sizes where $N \ge 64\ 000$
at $t = (n+\frac{1}{2}) \frac{L}{c_T}\ (n=0,1,2,\cdots )$ (the open arrows)
owing to the transverse sound waves traversing the whole periodic system.
Although the transient peaks encompass the whole time region. 
$S_4(k,t)$ is estimated at $t=t_4$, which maximizes the heterogeneity of the configuration overlap. 
We conjecture that the overall peak position can be identified for 
$N\le 640\ 00$ as indicated by the filled arrows. 
Figure \ref{fig:DH} (b) shows both of the structure factors for 2D SC as functions of wavenumber $k$.
While $S_4(k,t_4)$ exhibits strong divergence with $N$ at a small $k$, $S_B(k,t_B)$ exhibits no size dependence.
For 2D SC, Figs. \ref{fig:DH} (c) and (d) show that the $\alpha$-relaxation time $\tau_\alpha$, and
four-point time and length $t_4$ and $\xi_4$ exhibit strong size dependence owing to the vibration motion.
By contrast,  $t_B$ and $\xi_B$ exhibit no size dependence. 
For 3D KALJ,  $t_4$ and $t_B$ exhibit no size dependence for $N \ge 2500$.
The structure factors $S_4(k,t_4)$ and $S_B(k,t_B)$ in 
Fig. \ref{fig:DH} (e) show that there are no finite-size effects in the dynamics. 
For the entire range of temperatures under investigation, the two dynamic correlation lengths $\xi_4$ and $\xi_B$
exhibit a perfect coincidence, as shown in Fig. \ref{fig:DH} (f).

For 2D systems, these two correlation functions treat different aspects
 of dynamic fluctuation.
As described previously, the thermal amplitude $A_p = \sqrt{2\langle |\bvec{u}|^2 \rangle }$  
reaches a magnitude of $0.3\sigma_{11}$, or even exceeds this value
as the system becomes larger.
Most of the traditional correlation functions
for glassy dynamics,  including the standard MSD, self part of the intermediate 
scattering functions,  and four-point functions, are under the influence 
of such vibration motion. 
For a 2D system with small system sizes, 
the thermal fluctuation is weak enough that 
it does not mask the intermittent jump motion of individual particles
(giving rise to particle displacement with a magnitude on the order of $\sigma_{11}$)~\cite{Shiba2012}. 
The jump motion is further eclipsed by the vibrations with larger amplitudes
as the system size becomes larger, so the dual nature of the dynamics becomes indistinguishable. 
By contrast, the bond-breakage function is trivially free from coherent motion 
including long-wavelength vibrations because the broken-bond number  
assumes finite values only when particle rearrangement takes place.  
Therefore, the bond-breakage function takes over this role as a structural relaxation indicator. 
In line with this depiction,  the lack of size dependence of bond-breakage 
correlations ensures the existence of rearranging dynamics even in a large 2D system. 
It is worth mentioning that such a dual nature of DH in 2D system
is confirmed by the use of another dynamic correlation function 
based on a cage-relative variable~\cite{suppl}.

In conclusion, 2D supercooled liquids exhibit
strong thermal vibrational fluctuation whose 
amplitude grows infinitely in the limit of $N\rightarrow\infty$.
Our observations establish that the infinite fluctuations
are due to mechanism similar to that in 2D crystalline solids {\it i.e.}
enhancement of low-frequency sound waves.
In addition, the influence of such fluctuation  on the dynamic time and length is clarified.
The standard correlation functions fail to characterize the
original structural glassy relaxation, which can be retrieved
by quantifying the change in the relative particle positions.
The bond-breakage correlation function successfully
undertakes this role, and its correlation function exhibits
no system-size dependence. Therefore, the correlation
length $\xi_B$ is expected to represent the underlying length
inherent to the dynamics. It leads us to come to the second conclusion that
the modality of the glassy structural relaxation is
similar between the 2D and 3D systems after separating out
the 2D-specific thermal vibration. The infinitely growing
fluctuation could still affect the fundamental nature of a 2D
glass transition, but this is left for future investigation.

\begin{acknowledgments}
We thank Akira Onuki for motivating us to perform this work.
We also thank Kunimasa Miyazaki, Ryoichi Yamamoto, Atsushi Ikeda,
Hideyuki Mizuno, Masaki Oshikawa, Peter Keim, and Matthias Fuchs for helpful discussions.
This work was partially supported by
JSPS KAKENHI Grants No. 
JP25103010 (H. S.),
JP15H06263 and JP16H06018 (T. K.), and No. JP26400428 and No. JP16H00829 (K. K.).
H. S. and Y. Y. were also financially supported by the Building of Consortia for the 
Development of Human Resources in Science and Technology, MEXT, Japan.
The numerical calculations were carried out on SGI Altix ICE 8400EX
and XA at ISSP, University of Tokyo and on Fujitsu PRIMERGY RX300 S7 at
RCCS, NINS, Okazaki, Japan.
\end{acknowledgments}

\end{document}

%\preprint{APS/123-QED}
\title{Supplemental Material for  \\
Unveiling Dimensionality Dependence of Glassy Dynamics:   \\ 
2D Infinite Fluctuation Eclipses Inherent Structural Relaxation}

\author{Hayato Shiba}
%\email[E-mail: ]{shiba@imr.tohoku.ac.jp}
\affiliation{Institute for Solid State Physics, University of Tokyo, Chiba 277-8581, Japan}
\affiliation{Institute for Materials Research, Tohoku University, Sendai 980-8577, Japan}

\author{Yasunori Yamada}
%\altaffiliation[Present Address: ]{Beijing Computational Science Research Center, Beijing 100084, China}
\affiliation{Institute for Materials Research, Tohoku University, Sendai 980-8577, Japan}

\author{Takeshi Kawasaki}
\affiliation{Department of Physics, Nagoya University, Aichi 464-8602, Japan}
\author{Kang Kim}
\affiliation{Department of Physics, Niigata University, Niigata 950-2181, Japan}
\affiliation{Division of Chemical Engineering, Graduate School of Engineering Science, Osaka University, Osaka 560-8531, Japan}

\maketitle

\section{Simulation settings and models \label{sec:model} }
For a two-dimensional (2D) system, we employ
the binary 50:50 soft-core (SC)  potential given by 
\begin{equation}
v_{\alpha\beta} (r) = \epsilon_{\alpha\beta} \left( \frac{\sigma_{\alpha\beta} }{r}\right)^{12}
\end{equation}
for $r\le r_c = 2.21 \sigma_1$,  with $r_c$ as the cutoff length. 
The cubic smoothing function $v_{\alpha\beta} = B (a-r)^3 + C$ 
is applied for distances $r> r_c$, with $a, B,$ and $C$
satisfying continuity conditions at $r=r_c$ up to the second derivative of $v_{\alpha\beta}$.
The indices $\alpha ,\beta \in \{ 1,2 \}$ represent the particle species, 
where the particle number ratio is set to 50:50. 
The size and mass ratios are set to $\sigma_2 /\sigma_1 = 1.4$ and $m_2/m_1 = (\sigma_2 /\sigma_1)^2$, 
with $\sigma_{\alpha\beta} = (\sigma_\alpha + \sigma_\beta )/2$, 
and $\epsilon_{\alpha\beta} = \epsilon$ for all the pairs. 
The total number of particles $N$ is set to $N=250, 1000, 4000, 16\ 000, 64\ 000,$ and $256\ 000$,
and the box length $L$ is determined such that the number density $\phi = N/ L^2$ becomes 0.811. 
We simulated with a temperature $T=0.64$ in unit of $\epsilon / k_B$ %
and temperature dependence of the dynamic length is further investigated for $T=0.72,\ 0.80,\ 0.96,$ and $1.2$. 
Similar repulsive potential is employed in the previous
literature~\cite{Yamamoto1998a,Shiba2012}.

For the three-dimensional (3D) system,  we employ the well-known 
binary Kob-Andersen Lennard-Jones mixture~\cite{Kob1995} (KALJ) is used.  Its potential is given by 
\begin{equation}
v_{\alpha\beta}(r) = 4\epsilon_{\alpha\beta} \left[ \left( \frac{\sigma_{\alpha\beta}}{r} \right)^{12} - \left( \frac{\sigma_{\alpha\beta} }{r}\right)^6 \right].
\end{equation}
The energy and size ratios are given by $\epsilon_{12} /\epsilon_{11} = 1.5,\ \epsilon_{22}/\epsilon_{11} = 0.5$, and 
$\sigma_{12}/\sigma_{11} = 0.8,\ \sigma_{22}/\sigma_{11} = 0.88$, respectively. 
The particle number ratio between components 1 and 2 is set to 80:20.  
All the particles have the same mass $m$ {\it i.e.} $m_1 =m_2 =m$, and the interaction potentials 
are truncated at $r=2.5\sigma_{\alpha\beta}$.  
The total number of particles $N$ is set to $N=2\ 500, 20\ 000, 160\ 000,$ and $10\ 240\ 000$,
and the number density is fixed at $\phi = N/L^3 = 1.2$. 
We mainly simulate $T=0.47$ in unit of $\epsilon_{11}/k_B$, 
and temperature dependence of the dynamic length is also investigated
 for $T=0.5,\ 0.55,\ 0.6,\ 0.65,\ 0.7,$ and $0.75$ (See Fig. 4 (f) in the Letter). 

In the Letter, the simulation results (for both 2D and 3D systems)
are presented in terms of the reduced units
$\sigma_{11}$, $\epsilon_{11}/k_B$ for the length and temperature, with 
the Boltzmann constant assumed as unity ($k_B=1$).  
$\tau = \sqrt{m_1\sigma_{11}^2 / \epsilon_{11}}$ is employed as the unit of time. 
The simulations start from liquid state at $T=2.5$, 
and the system is allowed to relax at the target temperatures under the Langevin 
thermostat over a period  50 times longer  than the bond-breakage correlation time (mentioned later).
After then, the relaxation dynamics and the correlation functions are measured 
by using Newtonian dynamics simulations.

\section{Four-point and bond-breakage correlation functions}
In the Letter, the two different correlation functions are employed 
to  investigate the dynamic heterogeneity.  
The first four-point correlation function~\cite{Lacevic2003,Chandler2006,
Szamel2006,Flenner2010,Shiba2012,Kim2013a,Flenner2015}
characterize the mobility distribution by using the overlap function. 
It is a measure of the configuration overlap between two times separated by time $t$, 
\begin{equation}
W_j (t) = \Theta (a- |\bvec{r}_j(t) -\bvec{r}_j(0) | ),
\end{equation}
with $\Theta (x)$ being the Heaviside step function. 
In accordance with the standard practice, $a=0.3\sigma_{11}$ is employed.

The other bond-breakage correlation function~\cite{Shiba2012} 
concerns the coordination number fluctuation. 
In order to characterize particle rearrangements, 
taking place when a particle undergoing glassy relaxation escapes out of a cage, 
we measure how many neighbor particles exist (bond number) at the initial time
for each particle, and how many of them disappear (bond-breakage number) over a lapse of time $t$. 
The neighbor number is counted at  the initial time
\begin{equation}
r_{ij}(0) = |\bvec{r}_i(0) -\bvec{r}_j(0) | < b_1 \sigma_{\alpha\beta},
\end{equation}
with $r_{ij} (t)$ the distance between particles $i$ and $j$. 
We also count at a later time $t$ 
how many of these particles get separated around particle $i$ to satisfy 
\begin{equation}
r_{ij}(t) = |\bvec{r}_i(t) -\bvec{r}_j(t) |> b_2 \sigma_{\alpha\beta}.
\end{equation}
By using these criteria for the coordination number determination,
the bond-breakage number
is represented by a sum over its neighbors 
\begin{equation}
%Z_i(t) = \sum_{j=1}^N   X_{ij}(0) ( 1-Y_{ij}(t) ),
Z_i(t) = \sum_{j\in {\rm n.n.}}  \Theta (b_1\sigma_{\alpha\beta} -r_{ij}(t) )  \left[ 1 -  \Theta (b_2\sigma_{\alpha\beta} - r_{ij} (t)  ) \right]. 
\end{equation} 
For the threshold values, $(b_1,b_2) = (1.15,1.5)$ for 2D SC and $(1.3,1.65)$ for 3D KALJ 
are employed, so that contributions
from short-time reversible motion can be excluded. 

For both the functions, the dynamic susceptibilities are defined by
\begin{equation}
\begin{array}{ll} 
\chi_4 (t) = N [ \langle W(t)^2\rangle - \langle W(t)\rangle^2], & W(t) = N^{-1}\sum_{j=1}^N W_j(t)   \\
\chi_B(t) = N[ \langle Z(t)^2\rangle - \langle Z(t)\rangle^2 ], &  Z(t) = N^{-1} \sum_{j=1}^N Z_j(t). 
\end{array}
\end{equation}
For each function, the peak time $t_4$ and $t_B$ is defined as the peak time of these susceptibilities,
which can be defined for most of the cases (the exception in this Letter is 2D SC at $N=256\ 000$ 
because its $\chi_4(t)$ exhibits large-scale oscillation due to the acoustic vibration to smear out the peak). 
At the peak times, the dynamic heterogeneity is the most enhanced. 

To estimate the heterogeneity in the dynamics, 
the structure factor can be defined for each function: 
\begin{eqnarray}
S_4(k,t) = \frac{1}{N} \langle Q(\bvec{k},t) Q(-\bvec{k},t)\rangle, &\quad& Q(\bvec{k},t) = \sum_{j=1}^N W_j (t)  \exp [-i \bvec{k}\cdot\bvec{r}_j (0) ], \\
S_B (k,t) = \frac{1}{N} \langle P(\bvec{k},t) P(-\bvec{k},t) \rangle, &\quad& P(\bvec{k},t) = \sum_{j=1}^N Z_j(t) \exp [-i\bvec{k}\! \cdot\bvec{r}_j(0) ],
\end{eqnarray}
with the wavenumber $k=|\bvec{k}|$.

Because the heterogeneity characterized by $W(t)$ and $Z(t)$ becomes the most prominent at $t=t_4$ and 
$t=t_B$, respectively, we investigate length scale of the dynamics at these times. 
$S_4(k,t_4)$ and $S_B(k,t_B)$ are fitted to generalized Ornstein-Zernike (OZ) functions 
\begin{eqnarray}
S_4 (k, t_4) &=& \frac{S_{40}}{1+ [k \xi_4  ]^\alpha} \label{eq:OZ1} \\
S_B (k,t_B) &=& \frac{S_{B0}}{1+ [k \xi_B ]^\alpha},  \label{eq:OZ2}
\end{eqnarray}
where we employ $\alpha=2$ and $2.34$ for 2D SC and 3D KALJ, respectively.

\section{Cage-relative quantities}
In the Letter, dynamic heterogeneities are discussed in terms of four-point 
and bond-breakage correlation functions. 
In this section, we further characterize the dynamical heterogeneity for 2D SC, 
by using cage-relative quantities that are recently discussed in the
literature~\cite{2009Mazoyer,2011Mazoyer,2016Illing,Vivek2016}. 

The cage-relative displacement is given by 
\begin{equation}
\Delta \bvec{r}_{i,{\rm CR}}(t) = \Delta \bvec{r}_i (t)  - \frac{1}{N_{\rm n.n.}} \sum_{j\in {\rm n.n.}} \Delta \bvec{r}_j (t),
\end{equation}
with $\Delta\bvec{r}_j (t) = \bvec{r}_j (t)  - \bvec{r}_j (0)$ 
being the particle displacement 
and $N_{\rm n.n.}$ the neighbor number of the particle $i$. 
The summation in the second term is taken over all the neighbor particles $j$. 
The neighbors are defined as particles $j$ satisfying 
$r_{ij}(t) = |\bvec{r}_j (t) - \bvec{r}_j(0) | < 1.25\sigma_{\alpha\beta}$, 
where $\{\alpha ,\beta \} \in \{ 1,2 \}$ denote the particle species. 
This is for the ease of the simulations, and 
the possible standard way for defining the neighbors is to perform Vonoroi tessellation 
as reported in other papers~\cite{2009Mazoyer,2011Mazoyer}, which is not adopted
in the Letter.

The mean-square of the cage-relative displacement 
\begin{equation}
\langle | \Delta \bvec{r}_{\rm CR}(t) |^2 \rangle =  \left\langle \frac{1}{N} \sum_{i=1}^N |\Delta \bvec{r}_{i,{\rm CR}} (t) |^2 \right\rangle 
\end{equation}
is the cage-relative mean-square displacement (CR-MSD), which 
is shown for 2D SC in Fig. 1 (a) in the Letter.  
CR-MSD is an alternative of MSD $\langle | \Delta \bvec{r}(t) |^2 \rangle 
= \langle (1/N) \sum_i  |\Delta \bvec{r}_i (t) |^2 \rangle$  that 
considers only the relative motion of a particlea that is relative to the cage
This quantity is irrelevant to the coherent motion. 

To estimate the dynamic correlation length related to the cage-relative quantity,
we introduce ``cage-relative overlap function'' by extending the overlap function as 
\begin{equation}
D_j (t) = \Theta ( a - | \Delta \bvec{r}_{j,{\rm CR}} (t) | ), 
\end{equation}
with $a=0.3\sigma_{11}$.
Accordingly, the cage-relative dynamical susceptibility
$\chi_{\rm CR} (t) $ is defined by using the average overlap amount
of cage-relative configuration $D (t) = (1/N) \sum_{j=1}^N D_j (t)$ as 
\begin{equation}
\chi_{\rm CR} (t)  =  N [ \langle D(t)^2\rangle - \langle D(t)\rangle^2 ].
\end{equation}
In accord with the structure factors for four-point and bond-breakage correlation functions, 
the following structure factor $S_{\rm CR} (k,t)$  is calculated for its peak time $t=t_{\rm CR}$:
\begin{equation}
S_{\rm CR}(k,t) = \frac{1}{N} \langle R (\bvec{k},t) R(-\bvec{k},t) \rangle ,\qquad R(\bvec{k},t ) = \sum_{j=1}^N  D_j (t) \exp [-i\bvec{k}\cdot\bvec{r}_j (0) ]. 
\end{equation}
By fitting the structure factor $S_{\rm CR} (k,t_{\rm CR})$ to the Ornstein-Zernike function
\begin{equation}
S_{\rm CR} (k, t_{\rm CR}) = \frac{S_{{\rm CR}0}}{1+ [k \xi_{\rm CR}  ]^\alpha}\quad (\alpha =2) \label{eq:OZ3} 
\end{equation}
for various temperatures, we can estimate the temperature dependence of the cage-relative
dynamic length $\xi_{\rm CR}$.

\begin{figure}[b]
\includegraphics[height=0.36\linewidth]{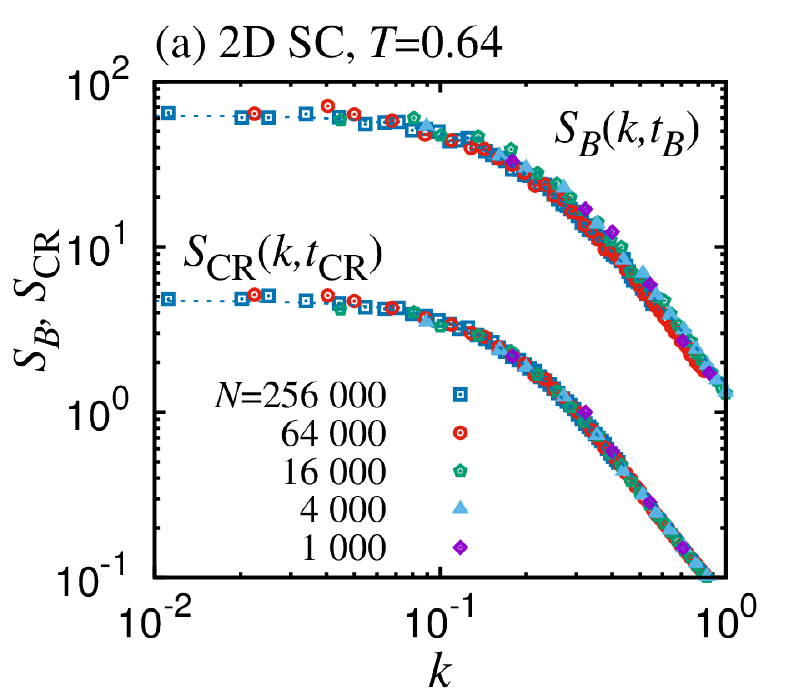}
\includegraphics[height=0.36\linewidth]{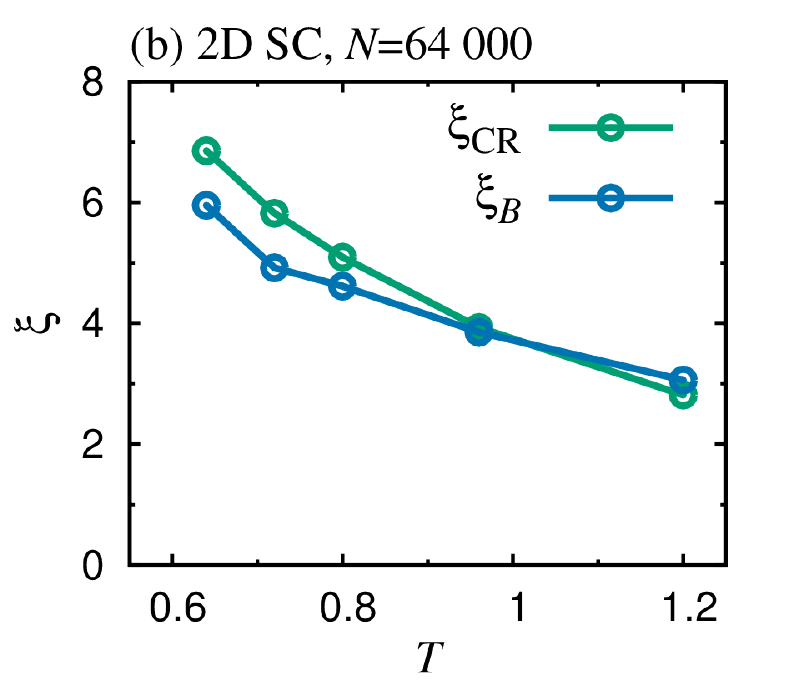}
\caption{\label{fig:cleng_2D} 
(a) System size dependence of structure factor
for the cage-relative overlap function $S_{\rm CR} (k,t_{\rm CR} )$ is plotted for 2D SC at $T=0.64$.   
The same data of bond-breakage structure factor $S_B (k,t_{B})$ as in Fig. 4 (b) is also plotted for the
sake of comparison.  Unlike the four-point structure factor $S_4 (k, t_4)$, neither of them exhibits
system size dependence.    (b)  Temperature dependence of the 
cage-relative and bond-breakage dynamic length $\xi_{\rm CR}$ and $\xi_B$ are 
plotted for 2D SC with $N=64\ 000$, indicating that dynamic heterogeneity
emerges similarly between the two correlation functions. 
}
\end{figure}

In Fig. \ref{fig:cleng_2D} (a), the structure factor  $S_{\rm CR}(k,t_{\rm CR})$ 
for 2D SC at $T=0.64$ is plotted together with $S_B(k,t_B)$ 
(the data are the same as in Fig. 4 (b) in the Letter). 
In Fig. \ref{fig:cleng_2D} (b), 
temperature dependence of  dynamic correlation length $\xi_{\rm CR}$  
for $N=64\ 000$  is compared with that of bond-breakage correlation length $\xi_B$.
The dynamical heterogeneity grows as the temperature is lowered, 
and the growth originates not in the coherent motion but in the cooperative rearrangement.
For a fixed temperature, the structure factors exhibit no system 
size dependence unlike the four-point structure factor $S_4(k,t_4)$ (see Fig. 4 (b) in the Letter). 
Therefore, the influence of the large-scale vibration that emerges similarly in four-point structure 
factors $S_4 (k,t)$ and dynamic lengths $\xi_4$ can also be removed by the use of the cage-relative quantities. 
The dynamic lengths in Fig. \ref{fig:cleng_2D} (b) exhibit good agreement with each other,
suggesting that the two correlation functions $S_B(k,t_B)$ and $S_{\rm CR}(k,t_{\rm CR})$ evaluates 
the same dynamical heterogeneity.

\newpage

%\bibliography{2015SKK}% Produces the bibliography via BibTeX.